\documentclass[11pt]{amsart}
\usepackage{hyperref}
\theoremstyle{definition}

\theoremstyle{remark}

\numberwithin{equation}{section}

\def \dd{{\rm d}}
\begin{document}

\title[Equilibrium positions of {\it n} wires]{Einstein's unified field theory
predicts the equilibrium
positions of {\it n} wires run by steady electric currents}%
\author{Salvatore Antoci}%
\address{Dipartimento di Fisica ``A. Volta'' and IPCF of CNR, Pavia, Italy}%
\email{Antoci@fisicavolta.unipv.it}%
\

\begin{abstract}
A particular exact solution of Einstein's Hermitian theory of
relativity is examined, after recalling that there is merit in
adding phenomenological sources to the theory, and in choosing the
metric like it was done long ago by Kur\c{s}uno\u{g}lu and H\'ely.
It is shown by intrinsic arguments, relying on the properties of
the chosen metric manifold, that the solution describes in
Einstein's theory the field of $n$ thin parallel wires at rest,
run by steady electric currents, and predicts their equilibrium
positions through the injunction that the metric must display
cylindrical symmetry in the infinitesimal neighbourhood of each
wire. In the weak field limit the equilibrium positions coincide
with the ones prescribed by Maxwell's electrodynamics.
\end{abstract}

\maketitle
\section{Introduction}\label{S1}
The theory of the nonsymmetric field, after an early attempt by
Einstein \cite{Einstein1925}, was separately developed in the same
years, but starting from different viewpoints, both by Einstein
\cite{ES1946, Einstein1948, EK1955} and by Schr\"odinger
\cite{Schroedinger1947a, Schroedinger1947b, Schroedinger1948,
Schroedinger1951}. Both of them thought that the theory had to be
some natural generalization of a successful predecessor, the
general theory of relativity of 1915. But, while Einstein decided
to deal with the nonsymmetric fundamental tensor and with the
nonsymmetric affine connection as independent entities,
Schr\"odinger's preference was for the purely affine approach.
Remarkably enough, they ended up with what soon appeared, from a
pragmatic standpoint, like two versions of one and the same
theory, because Schr\"odinger's ``final affine field laws'' just
look like the laws of Einstein's ``generalized theory of
gravitation'', to which a cosmological term is appended.\par It
was a conviction both by Einstein and by Schr\"odinger that, since
the theories had to be the completion of the theory of 1915,
neither a phenomenological energy tensor nor phenomenological
currents had to be added at the right-hand sides of the field
equations. However, exact solutions complying with this injunction
were never found: exact spherically symmetric solutions displayed
singularities \cite{Papapetrou1948,Wyman1950}, i.e. were useless
for the envisaged program. Approximate calculations by Callaway
\cite{Callaway1953}, although later shown to be incomplete by
Narlikar and Rao \cite{Narlikar Rao1956} and by Treder
\cite{Treder1957}, spread the conviction that the theory did not
contain the electromagnetic interaction; moreover, they too
allowed for singularities.  In any case, the formal simplicity of
the sets of equations proposed both by Einstein and by
Schr\"odinger did not find a counterpart in an equally simple and
satisfactory physical interpretation, and the interest aroused by
their endeavour began to fade. Already in the 1954/55 report to
the Dublin Institute for Advanced Studies, a disappointed Erwin
Schr\"odinger wrote: ``It is a disconcerting situation that ten
years endeavour of competent theorists has not yielded even a
plausible glimpse of Coulomb's law."\cite{Hittmair1987}.\par An
alternative viewpoint on those equations was however possible, in
some way bolstered by the ubiquitous presence of singularities in
the solutions found, by the very form of the contracted Bianchi
identities, and by the strict similarity of the new equations to
the field equations of 1915. It was expressed \cite{Hely1954a,
Hely1954b} in 1954 by H\'ely, who could avail in his attempt of
previous findings by Kur\c{s}uno\u{g}lu \cite{Kursunoglu1952a,
Kursunoglu1952b} and by Lichnerowicz \cite{Lichnerowicz1954} on
the choice of the metric. According to H\'ely, in the new theory
of Einstein phenomenological sources, in the form of a symmetric
energy tensor and of a conserved four-current, had to be appended
respectively at the right-hand sides of the field equations
(\ref{A4}) and (\ref{A5}), given in Appendix (\ref{A}). By
pursuing further H\'ely's proposal, and by relying on a precious
finding by Borchsenius \cite{Borchsenius1978}, the way for
appending sources to all the field equations while keeping the
choice of the metric done by H\'ely was later investigated
\cite{Antoci1991}, and is the subject of the next Section.

\section{Appending sources to Einstein's unified field
theory}\label{S2}
On a four-dimensional manifold, let
$\mathbf{g}^{ik}$ be a contravariant tensor density with an even
part $\mathbf{g}^{(ik)}$ and an alternating one
$\mathbf{g}^{[ik]}$:
\begin{equation}\label{2.1}
\mathbf{g}^{ik}=\mathbf{g}^{(ik)}+\mathbf{g}^{[ik]},
\end{equation}
and $W^i_{kl}$ be a general affine connection
\begin{equation}\label{2.2}
W^i_{kl}=W^i_{(kl)}+W^i_{[kl]}.
\end{equation}
The Riemann curvature tensor built from $W^i_{kl}$:
\begin{equation}\label{2.3}
R^i_{~klm}(W)=W^i_{kl,m}-W^i_{km,l}
-W^i_{al}W^a_{km}+W^i_{am}W^a_{kl},
\end{equation}
has two distinct contractions, $R_{ik}(W)=R^p_{~ikp}(W)$ and
$A_{ik}(W)=R^p_{~pik}(W)$ \cite{Schroedinger1950}. But the
transposed affine connection $\tilde{W}^i_{kl}=W^i_{lk}$ must be
considered too: from it, the Riemann curvature tensor
$R^i_{~klm}(\tilde{W})$ and its two contractions
$R_{ik}(\tilde{W})$ and $A_{ik}(\tilde{W})$ can be formed as well.
We aim at following the pattern of general relativity, which is
built from the Lagrangian density $\mathbf{g}^{ik}R_{ik}$, but now
any linear combination $\bar{R}_{ik}$ of the four above-mentioned
contractions is possible. A good choice \cite{Borchsenius1978},
for physical reasons that will become apparent later, is
\begin{equation}\label{2.4}
\bar{R}_{ik}(W)=R_{ik}(W)+\frac{1}{2}A_{ik}(\tilde{W}).
\end{equation}
Let us provisionally endow the theory with sources in the form of
a nonsymmetric tensor $P_{ik}$ and of a current density
$\mathbf{j}^i$, coupled to $\mathbf{g}^{ik}$ and to the vector
$W_i=W^l_{[il]}$ respectively. The Lagrangian density
\begin{equation}\label{2.5}
\mathbf{L}=\mathbf{g}^{ik}\bar{R}_{ik}(W)
-8\pi\mathbf{g}^{ik}P_{ik} +\frac{8\pi}{3}W_i\mathbf{j}^i
\end{equation}
is thus arrived at.  By performing independent variations of the
action $\int\mathbf{L}d\Omega$ with respect to $W^p_{qr}$ and to
$\mathbf{g}^{ik}$ with suitable boundary conditions we obtain the
field equations
\begin{eqnarray}\label{2.6}
-\mathbf{g}^{qr}_{~,p}+\delta^r_p\mathbf{g}^{(sq)}_{~,s}
-\mathbf{g}^{sr}W^q_{sp}-\mathbf{g}^{qs}W^r_{ps}\\\nonumber
+\delta^r_p\mathbf{g}^{st}W^q_{st} +\mathbf{g}^{qr}W^t_{pt}
=\frac{4\pi}{3}(\mathbf{j}^r\delta^q_p-\mathbf{j}^q\delta^r_p)
\end{eqnarray}
and
\begin{equation}\label{2.7}
\bar{R}_{ik}(W)=8\pi P_{ik}.
\end{equation}
By contracting eq. (\ref{2.6}) with respect to $q$ and $p$ we get
\begin{equation}\label{2.8}
\mathbf{g}^{[is]}_{~,s}={4\pi}\mathbf{j}^i.
\end{equation}
The very finding of this physically welcome equation entails
however that we cannot determine the affine connection $W^i_{kl}$
uniquely in terms of $\mathbf{g}^{ik}$: (\ref{2.6}) is invariant
under the projective transformation
${W'}^i_{kl}=W^i_{kl}+\delta^i_k\lambda_l$, where $\lambda_l$ is
an arbitrary vector field. Moreover eq. (\ref{2.7}) is invariant
under the transformation
\begin{equation}\label{2.9}
{W'}^i_{kl}=W^i_{kl}+\delta^i_k\mu_{,l}
\end{equation}
where $\mu$ is an arbitrary scalar. By following Schr\"odinger
\cite{Schroedinger1948,Schroedinger1950}, we write
\begin{equation}\label{2.10}
W^i_{kl}=\Gamma^i_{kl}-\frac{2}{3}\delta^i_kW_l,
\end{equation}
where $\Gamma^i_{kl}$ is another affine connection, by definition
constrained to yield $\Gamma^l_{[il]=0}$. Then eq. (\ref{2.6})
becomes
\begin{equation}\label{2.11}
\mathbf{g}^{qr}_{~,p}+\mathbf{g}^{sr}\Gamma^q_{sp}+\mathbf{g}^{qs}\Gamma^r_{ps}
-\mathbf{g}^{qr}\Gamma^t_{(pt)}
=\frac{4\pi}{3}(\mathbf{j}^q\delta^r_p-\mathbf{j}^r\delta^q_p)
\end{equation}
that allows one to determine $\Gamma^i_{kl}$ uniquely, under very
general conditions \cite{TH}, in terms of $\mathbf{g}^{ik}$. When
eq. (\ref{2.10}) is substituted in eq. (\ref{2.7}), the latter
comes to read
\begin{eqnarray}\label{2.12}
\bar{R}_{(ik)}(\Gamma)=8\pi P_{(ik)}\\\label{2.13}
\bar{R}_{[ik]}(\Gamma) =8\pi P_{[ik]}-\frac{1}{3}(W_{i,k}-W_{k,i})
\end{eqnarray}
after splitting the even and the alternating parts. Wherever the
source term is nonvanishing, a field equation loses its r\^ole,
and becomes a definition of some property of matter in terms of
geometrical entities; it is quite obvious that such a definition
must be unique. This occurs with eqs. (\ref{2.8}), (\ref{2.11})
and (\ref{2.12}), but it does not happen for eq. (\ref{2.13}).
This equation only prescribes that $\bar{R}_{[ik]}(\Gamma)-8\pi
P_{[ik]}$ is the curl of the arbitrary vector $W_i/3$; it is
therefore equivalent to the four equations
\begin{equation}\label{2.14}
\bar{R}_{[ik],l}(\Gamma)+\bar{R}_{[kl],i}(\Gamma)+\bar{R}_{[li],k}(\Gamma)
=8\pi\{P_{[ik],l}+P_{[kl],i}+P_{[li],k}\},
\end{equation}
that cannot specify $P_{[ik]}$ uniquely. We therefore scrap the
redundant tensor $P_{[ik]}$, like we scrapped the redundant affine
connection $W^i_{kl}$ of eq. (\ref{2.6}), and assume that matter
is described by the symmetric tensor $P_{(ik)}$, by the conserved
current density $\mathbf{j}^i$ and by the conserved current
\begin{equation}\label{2.15}
K_{ikl}=\frac{1}{8\pi}\{\bar{R}_{[ik],l}+\bar{R}_{[kl],i}+\bar{R}_{[li],k}\}.
\end{equation}
The general relativity of 1915, to which the present theory
reduces when $\mathbf{g}^{[ik]}=0$, suggests rewriting eq.
(\ref{2.12}) as
\begin{equation}\label{2.16}
\bar{R}_{(ik)}(\Gamma)=8\pi(T_{ik} -\frac{1}{2}s_{ik}s^{pq}T_{pq})
\end{equation}
where $s_{ik}=s_{ki}$ is the still unchosen metric tensor of the
theory, $s^{il}s_{kl}=\delta^i_k$, and the symmetric tensor
$T_{ik}$ will act as energy tensor.\par
     When sources are vanishing, equations (\ref{2.11}), (\ref{2.16}), (\ref{2.8}) and
(\ref{2.15}) reduce to the original equations of Einstein's
unified field theory, reported in Appendix (\ref{A}), because then
$\bar{R}_{ik}(\Gamma)$=${R}_{ik}(\Gamma)$; moreover they enjoy the
property of transposition invariance also when the sources are
nonvanishing. If $\mathbf{g}^{ik}$, $\Gamma^i_{kl}$,
$\bar{R}_{ik}(\Gamma)$ represent a solution with the sources
$T_{ik}$, $\mathbf{j}^i$ and $K_{ikl}$, the transposed quantities
$\tilde{\mathbf{g}}^{ik}=\mathbf{g}^{ki}$,
$\tilde{\Gamma}^i_{kl}=\Gamma^i_{lk}$ and
$\bar{R}_{ik}(\tilde{\Gamma})$= $\bar{R}_{ki}(\Gamma)$ represent
another solution, endowed with the sources $\tilde{T}_{ik}=T_{ik},
\tilde{\mathbf{j}}^i=-\mathbf{j}^i$ and
$\tilde{K}_{ikl}=-K_{ikl}$. Such a physically desirable outcome is
a consequence of the choice made \cite{Borchsenius1978} for
$\bar{R}_{ik}$. These equations intimate that Einstein's unified
field theory with sources should be interpreted like a
gravoelectrodynamics in a polarizable continuum, allowing for both
electric and magnetic currents. The study of the conservation
identities confirms the idea \cite{Antoci1991} and provides at the
same time the identification of the metric tensor $s_{ik}$. Let us
consider the invariant integral
\begin{equation}\label{2.17}
I=\int\left[\mathbf{g}^{ik}\bar{R}_{ik}(W)
+\frac{8\pi}{3}W_i\mathbf{j}^i\right]d\Omega.
\end{equation}
From it, when eq. (\ref{2.6}) is assumed to hold, by means of an
infinitesimal coordinate transformation we get the four identities
\begin{eqnarray}\label{2.18}
-(\mathbf{g}^{is}\bar{R}_{ik}(W)
+\mathbf{g}^{si}\bar{R}_{ki}(W))_{,s}
+\mathbf{g}^{pq}\bar{R}_{pq,k}(W)\\\nonumber
+\frac{8\pi}{3}\mathbf{j}^i(W_{i,k}-W_{k,i})=0.
\end{eqnarray}
This equation can be rewritten as
\begin{eqnarray}\label{2.19}
-2(\mathbf{g}^{(is)}\bar{R}_{(ik)}(\Gamma))_{,s}
+\mathbf{g}^{(pq)}\bar{R}_{(pq),k}(\Gamma)\\\nonumber
=2\mathbf{g}^{[is]}_{~,s}\bar{R}_{[ik]}(\Gamma) +\mathbf{g}^{[is]}
\left\{\bar{R}_{[ik],s}(\Gamma) +\bar{R}_{[ks],i}(\Gamma)
+\bar{R}_{[si],k}(\Gamma)\right\}
\end{eqnarray}
where the redundant variable $W^i_{kl}$ no longer appears. We
remind of eq. (\ref{2.16}) and, by following Kur\c{s}uno\u{g}lu
\cite{Kursunoglu1952a, Kursunoglu1952b} and H\'ely
\cite{Hely1954a, Hely1954b}, we assume that the metric tensor is
defined by the equation
\begin{equation}\label{2.20}
\sqrt{-s}s^{ik}=\mathbf{g}^{(ik)},
\end{equation}
where $s=\det{(s_{ik})}$; we shall use henceforth $s^{ik}$ and
$s_{ik}$ to raise and lower indices, $\sqrt{-s}$ to produce tensor
densities out of tensors. We define then
\begin{equation}\label{2.21}
\mathbf{T}^{ik}=\sqrt{-s}s^{ip}s^{kq}T_{pq}
\end{equation}
and the weak identities (\ref{2.19}), when all the field equations
hold, will take the form
\begin{equation}\label{2.22}
\mathbf{T}^{ls}_{;s}=\frac{1}{2}s^{lk}
(\mathbf{j}^i\bar{R}_{[ki]}(\Gamma) +K_{iks}\mathbf{g}^{[si]}),
\end{equation}
where the semicolon means covariant derivative with respect to the
Christoffel affine connection
\begin{equation}\label{2.23}
\left\{^{~i}_{k~l}\right\}
=\frac{1}{2}s^{im}(s_{mk,l}+s_{ml,k}-s_{kl,m})
\end{equation}
built with $s_{ik}$. The previous impression is strengthened by
eq. (\ref{2.22}): the theory, built in terms of a non-Riemannian
geometry, appears to entail a gravoelectrodynamics in a
dynamically polarized Riemannian spacetime, for which $s_{ik}$ is
the metric, where the two conserved currents ${\mathbf j}^i$ and
$K_{iks}$ are coupled \`a la Lorentz to $\bar{R}_{[ki]}$  and to
$\mathbf{g}^{[si]}$ respectively. Two versions of this
gravoelectrodynamics are possible, according to whether
$\mathbf{g}^{ik}$ is chosen to be either a real nonsymmetric or a
complex Hermitian tensor density. The constitutive relation
between electromagnetic inductions and fields is governed by the
field equations in a quite novel and subtle way: the link between
${\mathbf g}^{[ik]}$ and $\bar{R}_{[ik]}$ is not the simple
algebraic one usually attributed to the vacuum, with some metric
that raises or lowers indices, and builds densities from tensors.
It is a differential one, and a glance to the field equations
suffices to become convinced that understanding its properties is
impossible without first finding and perusing the exact solutions
of the theory.\par

This may seem a hopeless endeavour. However, a class of exact
solutions intrinsically depending on three coordinates has been
found; the method for obtaining them from vacuum solutions of the
general relativity of 1915 is described in Appendix (\ref{B}).
Some of these solutions happen to assume physical meaning when
source terms are appended to the field equations in the way
described in this Section. In particular, two static solutions
built in this way has been interpreted. One of them happens to
describe the general electrostatic field of $n$ localised charges
(\cite{ALM2005}) built by the four-current density ${\mathbf j}^i$
defined by equation (\ref{2.8}). As expected \cite{EI1949}, the
nonlinearity of the theory rules the singular behaviour of the
metric field $s_{ik}$ in the proximity of each charge. It rules it
in such a way that, with all the approximation needed to comply
with the experimental facts, the charges happen to be pointlike in
the metric sense and endowed with spherically symmetric
neighbourhoods only when they occupy mutual positions that
correspond to the ones dictated by Coulomb's law.\par

Another static, axially symmetric solution \cite{ALM2006, ALM2008}
displays instead $n$ aligned pole sources built with the
four-current $K_{ikl}$ defined by equation (\ref{2.15}). The
differential constitutive relation between ${\mathbf g}^{[ik]}$
and $\bar{R}_{[ik]}$, however, avoids the unphysical result that
these charges behave like magnetic monopoles would do, if they
were allowed for in the so-called Einstein-Maxwell theory. In
fact, the study of a particular solution endowed with three such
aligned charges shows that these pointlike charges interact with
forces not depending on their mutual distance. In the Hermitian
version of the theory, charges with opposite signs happen to
mutually attract, hence they are permanently confined entities,
like it was already shown by Treder \cite{Treder1957} with
approximate calculations based both on the E.I.H. \cite{EIH1938,
EI1949} and on the Papapetrou \cite{Papapetrou1951} method.\par In
the present paper the behaviour of a solution displaying $n$
steady currents built with ${\mathbf j}^i$ and running on parallel
wires is instead investigated, by using $s_{ik}$ as metric tensor.

\section{The equilibrium conditions of
steady electric currents running on $n$ parallel wires}\label{S3}

This solution belongs to the class described in Appendix
(\ref{B}); in it, only one of equations (\ref{2.8}) is not
trivially satisfied, and reads
\begin{equation}\label{3.1}
{\mathbf g}^{[3s]}_{~,s}=
i\left(\sqrt{-h}h^{\varrho\sigma}\xi_{,\sigma}\right)_{,\varrho}
=4\pi{\mathbf j}^3(x^{\lambda}).
\end{equation}
Like the electrostatic solution, this one too is obtained by
assuming that $h_{ik}$ has the Minkowski form
\begin{equation}\label{3.2}
h_{ik}={\rm diag}{(-1, -1, -1, 1)},
\end{equation}
with respect to the coordinates $x^1=x$, $x^2=y$, $x^3=z$,
$x^4=t$. In these coordinates its fundamental form $g_{ik}$,
defined by (\ref{B2}), reads:
\begin{equation}\label{3.3}
g_{ik}=\left(\begin{array}{rrrr}
 -1 &  0 &  e & 0 \\
  0 & -1 &  f & 0 \\
 -e & -f &  v & c \\
  0 &  0 & -c & 1
\end{array}\right),
\end{equation}
with
\begin{equation}\label{3.4}
v=-1-c^2+e^2+f^2
\end{equation}
and
\begin{equation}\label{3.5}
e=i\xi_{,x}, \ f=i\xi_{,y}, \ c=-i\xi_{,t}, \ \ i=\sqrt{-1}, \ \
\xi_{,xx}+\xi_{,yy}-\xi_{,tt}=0.
\end{equation}
Let us consider the particular, static solution for which
\begin{equation}\label{3.6}
\xi= \sum_{k=1}^n l_k \ln q_k
\end{equation}
where
\begin{equation}\label{3.7}
q_k=\left[(x-x_k)^2+(y-y_k)^2\right]^{1/2},
\end{equation}
and $l_k$, $x_k$, $y_k$ are arbitrary real constants. Then one
finds
\begin{eqnarray}\label{3.8}
e=i\sum_{k=1}^n l_k\frac{x-x_k}{q_k^2},  \ f=i\sum_{k=1}^n
l_k\frac{y-y_k}{q_k^2}, \ c=0,\\\label{3.9}
v=-1-\sum_{k=1}^n\frac{l_k^2}{q_k^2}\\\nonumber
-\left[\sum_{k,k'=1 }^n l_k
l_{k'}\frac{(x-x_k)(x-x_{k'})+(y-y_k)(y-y_{k'})}{q_k^2
q_{k'}^2}\right]_{k\ne k'}.
\end{eqnarray}
We note in passing that, despite its r\^ole as component of the
fundamental tensor, $v=-1+\frac12g_{[ik]}g^{[ik]}$ is an invariant
quantity.\par In this solution the vacuum field equation ${\mathbf
g}^{[is]}_{~,s}=0$ is satisfied everywhere, with the exception of
the positions $x=x_k$, $y=y_k$, $k=1,..,n$, of the wires in the
representative space, while, due to the additional conditions
(\ref{B3}), the additional invariant equation
\begin{equation}\label{3.10}
g_{[ik],l}+g_{[kl],i}+g_{[li],k}=0,
\end{equation}
is fulfilled everywhere. As far as the skew fields are concerned,
we are there\-fore inclined to interpret physically this solution
as the field produced by $n$ steady electric currents running
along thin wires drawn parallel to the $z$ coordinate axis. But in
order to do so, we need the proof, coming from the symmetric field
$s_{ik}$, that the wires are indeed thin in the metric sense, and
that the metric is endowed with cylindrical symmetry in the
infinitesimal neighbourhood of each wire. The latter property is
required for the wires to be in static equilibrium, in keeping
with a deep intuition present in \cite{EI1949}. To provide this
proof, we shall examine the square $\dd s^2$ of the interval,
defined by (\ref{B13}), that in the present case happens to read
\begin{equation}\label{3.11}
\dd s^2=\sqrt{-v}(\dd t^2-\dd x^2-\dd y^2- \dd z^2)+\frac{(\dd
\xi)^2}{\sqrt{-v}}.
\end{equation}
The first term of the interval is conformally flat. Due to the
existence of two Killing vectors, respectively along $t$ and along
$z$, and both orthogonal to each $x,y$ two-surface, it suffices
that we examine the problem on a given $x,y$ two-surface of the
manifold. We shall prove that the sections of the wires by the
given two-surface are pointlike in the metrical sense, and we
shall require that $s_{ik}$, in the infinitesimal neighbourhood of
each point $x=x_k$, $y=y_k$ of that two-surface, is endowed with
invariance under rotation around these points.\par The first
question is soon answered. In fact, from the behaviour of $v$,
defined by (\ref{3.9}), in an infinitesimal neighbourhood (in the
``Bildraum'' sense) of $x=x_k$, $y=y_k$, one gathers that the
first term
\begin{equation}\label{3.12}
\dd s_1^2=\sqrt{-v}(\dd t^2-\dd x^2-\dd y^2- \dd z^2)
\end{equation}
will vanish like $q_k$ when $q_k\rightarrow 0$. The second term
\begin{equation}\label{3.13}
\dd s_2^2=\frac{(\dd \xi)^2}{\sqrt{-v}}
\end{equation}
is now examined in the chosen neighbourhood. Due to the definition
(\ref{3.6}) of $\xi$, the numerator $(\dd \xi)^2$ will keep there
a finite value, while the denominator $\sqrt{-v}$ diverges like
$q_k^{-1}$ when $q_k\rightarrow 0$. Therefore the term
(\ref{3.13}) too shall vanish like $q_k$ in the considered,
infinitesimal neighbourhood. One concludes that, when $s_{ik}$ is
the metric tensor, the $n$ parallel wires of this solution will be
infinitely thin in the metric sense for any physically reasonable
choice of their mutual positions.\par We examine now the second
question, whether and under what conditions the metric field
$s_{ik}$ will exhibit rotational symmetry in the infinitesimal
neighbourhood of each point $x=x_k$, $y=y_k$ of the considered
two-surface. Let us imagine approaching the $k$-th wire along the
line defined by
\begin{equation}\label{3.14}
x-x_k=n_xq_k, \ \ y-y_k=n_yq_k,
\end{equation}
where $n_x$ and $n_y$ are constants for which $n_x^2+n_y^2=1$, but
otherwise arbitrary. To study $\dd s_1^2$ we need evaluating
$\sqrt{-v}$ in the infinitesimal neighbourhood of $x=x_k$,
$y=y_k$. Let us call this quantity
$\left\{\sqrt{-v}\right\}_{(k)}$. One first extracts from the root
the common factor $l_k/q_k$, and then subjects the other factor to
a Taylor's expansion truncated to terms that vanish like $q_k$
when $q_k\rightarrow 0$. Higher order terms will not influence the
final result. One then writes:
\begin{equation}\label{3.15}
\left\{\sqrt{-v}\right\}_{(k)}\simeq \frac{l_k}{q_k}
\left[1+\frac{q_k}{2l_k}\sum_{k'\ne k}^n
l_{k'}\frac{n_x(x_k-x_{k'})+n_y(y_k-y_{k'})}{d^2_{kk'}}\right],
\end{equation}
where
\begin{equation}\label{3.16}
d^2_{kk'}=(x_k-x_{k'})^2+(y_k-y_{k'})^2.
\end{equation}
If the term with the summation symbol were lacking,
$\left\{\sqrt{-v}\right\}_{(k)}$ would display rotational
symmetry, in the ``Bildraum'' sense, in the infinitesimal
neighbourhood of $x=x_k$, $y=y_k$. Since $n_x$, $n_y$ fulfill
$n_x^2+n_y^2=1$, but are otherwise arbitrary, the mentioned
symmetry only occurs when the conditions
\begin{equation}\label{3.17}
\sum_{k'\ne k}^n l_{k'}\frac{x_k-x_{k'}}{d^2_{kk'}}=0, \ \
\sum_{k'\ne k}^n l_{k'}\frac{y_k-y_{k'}}{d^2_{kk'}}=0,
\end{equation}
are severally satisfied. When this occurs the interval $\dd
s_1^2$, given by (\ref{3.12}), will be endowed with rotational
symmetry in an infinitesimal neighbourhood surrounding $x=x_k$,
$y=y_k$ in an intrinsic, geometric sense.\par We examine now $\dd
s_2^2$, defined by (\ref{3.13}), in the infinitesimal
neighbourhood of $x=x_k$, $y=y_k$ of the considered $x,y$
two-surface. Since $\xi$ is defined by (\ref{3.6}), one can write
\begin{eqnarray}\label{3.18}
\left\{\dd \xi\right\}_{(k)}=\left\{\xi_{,x}\dd x +\xi_{,y}\dd
y\right\}_{(k)}\\\nonumber\simeq\left[\frac{l_k
n_x}{q_k}+\sum_{k'\ne k}^n
l_{k'}\frac{x_k-x_{k'}}{d^2_{kk'}}\right]\dd x +\left[\frac{l_k
n_y}{q_k}+\sum_{k'\ne k}^n
l_{k'}\frac{y_k-y_{k'}}{d^2_{kk'}}\right]\dd y,
\end{eqnarray}
by neglecting all the terms that vanish when $q_k\rightarrow 0$,
because they will not influence  the final result. To calculate
$\left\{(\sqrt{-v})^{-1}\right\}_{(k)}$, let us extract from
$(\sqrt{-v})^{-1}$ the factor $q_k/l_k$, and then expand the other
factor in Taylor's series around $x=x_k$, $y=y_k$. We truncate the
expansion at the term linear in $q_k$, because higher order terms
do not influence the final outcome. Therefore we write
\begin{equation}\label{3.19}
\left\{(\sqrt{-v})^{-1}\right\}_{(k)}\simeq \frac{q_k}{l_k}
\left[1-\frac{q_k}{2l_k}\sum_{k'\ne k}^n
l_{k'}\frac{n_x(x_k-x_{k'})+n_y(y_k-y_{k'})}{d^2_{kk'}}\right],
\end{equation}
and calculate $\left\{(\dd\xi)^2/\sqrt{-v}\right\} _{(k)}$ from
(\ref{3.18}) and (\ref{3.19}). One finds that $\dd s_2^2$ in
general does not exhibit rotational symmetry in the infinitesimal
neighbourhood of $x=x_k$, $y=y_k$. Only if the conditions
(\ref{3.17}) are imposed, $\left\{(\dd\xi)^2/\sqrt{-v}\right\}
_{(k)}$ comes to read
\begin{equation}\label{3.20}
\left\{(\dd\xi)^2/\sqrt{-v}\right\} _{(k)}\simeq\frac{l_k}{q_k}
\left(n_x^2\dd x^2+n_y^2\dd y^2+2n_xn_y\dd x\dd y\right),
\end{equation}
i.e. it defines a two-dimensional interval endowed with rotational
symmetry in the intrinsic, geometric sense. Therefore, only if the
conditions (\ref{3.17}) are imposed, both $\dd s_1^2$ and $\dd
s_2^2$ become endowed with rotational symmetry, in an intrinsic,
geometric sense, in the infinitesimal neighbourhoods around  each
one of the points $x=x_k$, $y=y_k$, and the same property will be
exhibited by the interval $\dd s^2$, defined by (\ref{3.11}).\par
\section{Conclusion}\label{S4}
The equations (\ref{3.1}) together with (\ref{3.8}), and
(\ref{3.10}), as well as the general formulation of Section
(\ref{S2}), already led to think that the present solution
physically describes, in Einstein's unified field theory, the
field originated by steady electric currents running on $n$
parallel wires. The scrutiny of the interval (\ref{3.11}) confirms
this interpretation. In fact, when the metric is $s_{ik}$, the
parallel wires of the ``Bildraum'' turn out to be infinitely thin
parallel wires in the metric sense. Moreover, the infinitesimal
neighbourhood of each of these wires happens to be endowed with
cylindrical symmetry only when the conditions (\ref{3.17}), with
their distinct flavour of {\it d\`ej\'a vu}, are satisfied. The
equilibrium conditions for $n$ parallel wires run by steady
currents in Maxwell's electrodynamics are just written in that
way. It is true that we shall not be deceived by retrieving their
exact replica in Einstein's unified field theory with sources,
because this is just an accidental occurrence due to the
particular coordinates adopted when solving the field equations.
When measured along geodesics built with the metric $s_{ik}$,
distances and angles are different from the ones that would
prevail if the conditions (\ref{3.17}) were read as if they would
hold in a Minkowski metric. But there is no doubt that in the weak
field limit the particular exact solutions of both theories
describe one and the same physical reality.\par As stressed long
ago by Einstein in the Introduction of both (\cite{EIH1938}) and
(\cite{EI1949}), there is one distinct advantage in working with
such nonlinear theories as the general relativity of 1915, or its
nonsymmetric generalization. While in the linear physics of, say,
Maxwell's electrodynamics, the field equations and the equations
of motion need to be separately postulated, this is no longer the
case in, e.g., the Hermitian theory with sources. In the latter
theory, all what is needed is solving the field equations. From
the very solution one learns the equations of motion, by just
imposing symmetry conditions on the metric around the
singularities that are used to represent the physical objects. In
the case of a static manifold, one learns the equilibrium
conditions of such objects, like it has been shown, once more,
through the exact solution of the previous Section.

\appendix
\section{Hermitian field equations without sources}\label{A}
We consider here Einstein's unified field theory in the Hermitian
version (\cite{Einstein1948}). A given geometric quantity
\cite{Schouten1954} will be called hereafter Hermitian with
respect to the indices $i$ and $k$, both either covariant or
contravariant, if the part of the quantity that is symmetric with
respect to $i$ and $k$ is real, while the part that is
antisymmetric is purely imaginary. Let us consider the Hermitian
fundamental form $g_{ik}=g_{(ik)}+g_{[ik]}$ and the affine
connection $\Gamma^i_{kl}=\Gamma^i_{(kl)}+\Gamma^i_{[kl]}$,
Hermitian with respect to the lower indices; both entities depend
on the real coordinates $x^i$, with $i$ running from 1 to 4. We
define also the Hermitian contravariant tensor $g^{ik}$ by the
relation
\begin{equation}\label{A1}
g^{il}g_{kl}=\delta^i_k,
\end{equation}
and the contravariant tensor density
$\mathbf{ g}^{ik}=(-g)^{1/2}g^{ik}$, where $g\equiv\text{det} (g_{ik})$
is a real quantity. Then the field equations of Einstein's unified field
theory in the complex Hermitian form \cite{Einstein1948} read
\begin{eqnarray}\label{A2}
g_{ik,l}-g_{nk}\Gamma^n_{il}-g_{in}\Gamma^n_{lk}=0,\\\label{A3}
\mathbf{ g}^{[is]}_{~~,s}=0,\\\label{A4}
R_{(ik)}(\Gamma)=0,\\\label{A5}
R_{[ik],l}(\Gamma)+R_{[kl],i}(\Gamma)+R_{[li],k}(\Gamma)=0;
\end{eqnarray}
$R_{ik}(\Gamma)$ is the Hermitian Ricci tensor
\begin{equation}\label{A6}
R_{ik}(\Gamma)=\Gamma^a_{ik,a}-\Gamma^a_{ia,k}
-\Gamma^a_{ib}\Gamma^b_{ak}+\Gamma^a_{ik}\Gamma^b_{ab}.
\end{equation}

\section{Solutions depending on three coordinates \cite{Antoci1987a}}\label{B}
 We assume that Greek indices take the values 1,2 and 4,
while Latin indices run from 1 to 4. Let the real symmetric tensor
$h_{ik}$ be the metric for a vacuum solution to the field
equations of the general relativity of 1915, which depends on the
three co-ordinates $x^{\lambda}$, not necessarily all spatial in
character, and for which $h_{\lambda 3}=0$. We consider also an
antisymmetric purely imaginary tensor $a_{ik}$, which depends too
only on the co-ordinates  $x^{\lambda}$, and we assume that its
only nonvanishing components are $a_{\mu 3}=-a_{3 \mu}$. Then we
form the mixed tensor
\begin{equation}\label{B1}
\alpha_i^{~k}=a_{il}h^{kl}=-\alpha^k_{~i},
\end{equation}
where $h^{ik}$ is the inverse of $h_{ik}$, and we define the
Hermitian fundamental form $g_{ik}$ as follows:
\begin{eqnarray}\nonumber
g_{\lambda\mu}=h_{\lambda\mu},\\\label{B2}
g_{3\mu}=\alpha_3^{~\nu}h_{\mu\nu},\\\nonumber
g_{33}=h_{33}-\alpha_3^{~\mu}\alpha_3^{~\nu}h_{\mu\nu}.
\end{eqnarray}
When the three additional conditions
\begin{equation}\label{B3}
\alpha^3_{~\mu,\lambda}-\alpha^3_{~\lambda,\mu}=0
\end{equation}
are fulfilled, the affine connection $\Gamma^i_{kl}$ which solves
eqs. (\ref{A2}) has the nonvanishing components
\begin{eqnarray}\label{B4}
\Gamma^{\lambda}_{(\mu\nu)}=\left\{^{~\lambda}_{\mu~\nu}\right\}_{(h)},
\\\nonumber
\Gamma^{\lambda}_{[3\nu]}=\alpha^{~\lambda}_{3~,\nu}
-\left\{^{~3}_{3~\nu}\right\}_{(h)}\alpha^{~\lambda}_3
+\left\{^{~\lambda}_{\rho~\nu}\right\}_{(h)}\alpha^{~\rho}_3,
\\\nonumber
\Gamma^3_{(3\nu)}=\left\{^{~3}_{3~\nu}\right\}_{(h)},
\\\nonumber
\Gamma^{\lambda}_{33}=\left\{^{~\lambda}_{3~3}\right\}_{(h)}
-\alpha^{~\nu}_3\left(\Gamma^{\lambda}_{[3\nu]}
-\alpha^{~\lambda}_3\Gamma^3_{(3\nu)}\right);
\end{eqnarray}
we indicate with $\left\{^{~i}_{k~l}\right\}_{(h)}$ the
Christoffel connection built with $h_{ik}$. We form now the Ricci
tensor (\ref{A6}). When eqs. (\ref{A3}), i.e., in our case, the
single equation
\begin{equation}\label{B5}
(\sqrt{-h}~\alpha^{~\lambda}_3 h^{33})_{,\lambda}=0,
\end{equation}
and the additional conditions, expressed by eqs. (\ref{B3}), are
satisfied, the components of $R_{ik}(\Gamma)$ can be written as
\begin{eqnarray}\nonumber
R_{\lambda\mu}=H_{\lambda\mu},
\\\label{B6}
R_{3\mu}=\alpha^{~\nu}_3H_{\mu\nu}+\left(\alpha^{~\nu}_3
\left\{^{~3}_{3~\nu}\right\}_{(h)}\right)_{,\mu},
\\\nonumber
R_{33}=H_{33}-\alpha^{~\mu}_3\alpha^{~\nu}_3H_{\mu\nu},
\end{eqnarray}
where $H_{ik}$ is the Ricci tensor built with
$\left\{^{~i}_{k~l}\right\}_{(h)}$. $H_{ik}$ is zero when $h_{ik}$
is a vacuum solution of the field equations of general relativity,
as supposed; therefore, when eqs. (\ref{B3}) and (\ref{B5}) hold,
the Ricci tensor, defined by eqs. (\ref{B6}), satisfies eqs.
(\ref{A4}) and (\ref{A5}) of the Hermitian theory of
relativity.\par The task of solving equations
(\ref{A2})-(\ref{A5}) reduces, under the circumstances considered
here, to the simpler task of solving eqs. (\ref{B3}) and
(\ref{B5}) for a given $h_{ik}$.\footnote{This method of solution
obviously applies to Schr\"odin\-ger's purely affine theory
\cite{Schroedinger1951} too.}\par Let us assume, like in Section
(\ref{S2}), that the metric tensor is defined by the equation
\begin{equation}\label{B7}
\sqrt{-s}s^{ik}=\mathbf{g}^{(ik)},
\end{equation}
where $s^{il}s_{kl}=\delta^i_k$ and $s=\det{(s_{ik})}$. When the
fundamental tensor $g_{ik}$ has the form (\ref{B2}) it is
\begin{equation}\label{B8}
\sqrt{-g}=\sqrt{-h},
\end{equation}
where $h\equiv\det(h_{ik})$, and
\begin{equation}\label{B9}
\det{\left(g^{(ik)}\right)}=\frac{1-g^{3\tau}g_{3\tau}}{h}.
\end{equation}
Therefore
\begin{equation}\label{B10}
\sqrt{-s}=\sqrt{-h}\left(1-g^{3\tau}g_{3\tau}\right)^{1/2},
\end{equation}
hence
\begin{equation}\label{B11}
s^{ik}=g^{(ik)}\left(1-g^{3\tau}g_{3\tau}\right)^{-1/2}.
\end{equation}
The nonvanishing components of $s_{ik}$ then read
\begin{eqnarray}\nonumber
s_{\lambda\mu}=\left(1-g^{3\tau}g_{3\tau}\right)^{1/2}h_{\lambda\mu}
+\left(1-g^{3\tau}g_{3\tau}\right)^{-1/2}
h_{33}\alpha^3_{~\lambda}\alpha^3_{~\mu},\\\label{B12}
s_{33}=\left(1-g^{3\tau}g_{3\tau}\right)^{1/2}h_{33},
\end{eqnarray}
and the square of the interval $\dd s^2=s_{ik}\dd x^i\dd x^k$
eventually comes to read
\begin{equation}\label{B13}
\dd s^2=\left(1-g^{3\tau}g_{3\tau}\right)^{1/2}h_{ik}\dd x^i\dd
x^k
-\left(1-g^{3\tau}g_{3\tau}\right)^{-1/2}h_{33}\left(\dd\xi\right)^2.
\end{equation}
In keeping with (\ref{B3}), we have defined $\alpha^3_{~\mu}$ as
\begin{equation}\label{B14}
\alpha^3_{~\mu}=i\xi_{,\mu},
\end{equation}
in terms of the real function $\xi(x^{\lambda})$.

\newpage

\bibliographystyle{amsplain}

\end{document}